\documentclass[%
aps,
pra,
superscriptaddress,
tightenlines,
showpacs,showkeys,
a4paper,
12pt,
notitlepage
]{revtex4-1}
\usepackage[english]{babel}
\usepackage{amssymb,amsmath,stmaryrd,array}

\usepackage{graphicx}
\usepackage{subfig}

\makeatletter

\newcommand{\sca}[2]{\ensuremath{\bigl({#1}\cdot{#2}\bigr)}}

\newcommand{\hcnj}[1]{{#1}^{\dagger}}
\newcommand{\tcnj}[1]{{#1}^{T}}

\newcommand{\pdrs}[1]{\partial_{#1}}

\newcommand{\diag}{\mathop{\rm diag}\nolimits}

\newcommand{\adj}{\mathop{\rm adj}\nolimits}

\newcommand{\mum}{$\mu$m}

 \newcommand{\bs}[1]{\boldsymbol{#1}}
 \newcommand{\vc}[1]{\mathbf{#1}}
 \newcommand{\mvc}[1]{\mathbf{#1}}
 \newcommand{\uvc}[1]{\hat{\mathbf{#1}}}
 
 \newcommand{\ind}[1]{\mathrm{#1}}

\newcommand{\rw}{\mathrm{RW}}

\newcommand{\inc}{\mathrm{inc}}
\newcommand{\refl}{\mathrm{refl}}
\newcommand{\transm}{\mathrm{trm}}
\newcommand{\vac}{\mathrm{vac}}
\newcommand{\med}{\mathrm{m}}

\makeatother

\begin{document}
\DeclareGraphicsExtensions{.jpg,.png,.pdf}
\title{Light-induced pitch transitions in photosensitive cholesteric liquid
  crystals: Effects of anchoring energy }

\author{Tetiana~N.~Orlova}
\email[Email address: ]{orlovat@gmail.com}
\affiliation{%
 Institute of Physics of National Academy of Sciences of Ukraine,
 prospekt Nauki 46,
 03680 Kiev, Ukraine} 
\affiliation{%
LOMA (Laboratoire Ondes et Matiere d'Aquitaine), University Bordeaux 1,
351 Cours de la Liberation, 33400 Talence, France
}

\author{Roman~I.~Iegorov}
\email[Email address: ]{rommel.ua@gmail.com}
\affiliation{%
 Institute of Physics of National Academy of Sciences of Ukraine,
 prospekt Nauki 46,
 03680 Kiev, Ukraine} 
\affiliation{%
UFOLAB (Ultrafast Optics \& Lasers Laboratory), Bilkent University,
06800 Bilkent, Ankara, Turkey
}

\author{Alexei~D.~Kiselev}
\email[Email address: ]{kiselev@iop.kiev.ua}
\affiliation{%
 Institute of Physics of National Academy of Sciences of Ukraine,
 prospekt Nauki 46,
 03680 Kiev, Ukraine}

\date{\today}

\begin{abstract}
 We experimentally study how the cholesteric
pitch, $P$, depends on the equilibrium pitch $P_0$
in planar liquid crystal (LC) cells with both strong and
semistrong anchoring conditions. 
The cholesteric phase was induced by
dissolution in the nematic LC the right-handed chiral dopant 
7-dehydrocholesterol (7-DHC, provitamin $D_3$) which 
transforms to left-handed tachysterol 
under the action of uv irradiation at the wavelength of $254$~nm. 
By using the model of photoreaction kinetics we obtain
the dependencies of isomer concentrations and, therefore, 
of the equilibrium pitch on the uv irradiation dose. 
The cholesteric pitch was measured  
as a function of irradiation time
using the polarimetry method.
In this method, the pitch is estimated from
the experimental data on the irradiation time dependence 
of the ellipticity of light transmitted through the LC cells. 
It is found that the resulting dependence of the twist parameter
$2 D/P$ ($D$ is the cell thickness)
on the free twisting number parameter $2 D/P_0$ shows jump-like behavior
and agrees well with the known theoretical results  
for the anchoring potential of Rapini-Papoular form.
\end{abstract}

\pacs{%
61.30.Hn, 64.70.M-, 42.70.Df, 42.70.Gi
}
\keywords{%
cholesteric liquid crystal; helix pitch; anchoring energy, polarization of light 
} 
 \maketitle

\section{Introduction}
\label{sec:intro}

In equilibrium structures of chiral nematic liquid crystals
also known as \textit{cholesteric liquid crystals}
(CLC)  molecules
align on average along a local unit director $\uvc{n}(\vc{r})$ that
rotates in a helical fashion about a uniform twist
axis~\cite{Gennes:bk:1993}.  
This tendency of CLCs to form helical twisting patterns
is caused by the presence of anisotropic molecules with
no mirror plane~---~so-called chiral molecules
(see~\cite{Harr:rmp:1999,Kitzerow:bk:2001} for reviews).

In planar CLC cells bounded by two parallel substrates
orientational structures
(director configurations) are strongly affected by the
anchoring conditions at the boundary surfaces.
These conditions break the translational symmetry along the twisting
axis and, in general, the helical form of the director field will be distorted.

Nevertheless, when the anchoring conditions are planar and
out-of-plane deviations of the director are suppressed, it might
be expected that the configurations still have the form of the
ideal helical structure: 
\begin{align}
\label{eq:director}
\uvc{n}=\cos\phi\,\uvc{x}+\sin\phi\,\uvc{y}, 
\quad
\phi= q z+\phi_0,
\end{align}
where $q=2 \pi/P$ is the \textit{helix wave number}
and $\phi_0$ is the phase at $z=0$.
But, by contrast with  the case of
unbounded CLCs, the helix twist wave number $q$ will now differ
from $q_0=2\pi/P_0$.

A mismatch between 
the twist imposed by the boundary conditions
and the equilibrium pitch $P_0$ may 
produce two
metastable twisting states that are degenerate in energy and can be
switched either way by applying an electric
field~\cite{Berrem:jap:1981}.  This bistability underlines the mode of
operation of bistable liquid crystal 
devices~\cite{Xie:jap:1998,Zhuang:apl:1999,Xie:jap2:2000,Kwok:apl:2003}.

More generally the metastable twisting states in CLC cells appear
as a result of interplay between the bulk and the surface
contributions to the free energy. 
The free twisting number $q_0$ and the anchoring energy
are among  
the key factors that govern their properties. 
Specifically, varying $q_0$ will
change the twist wave number of the twisting state, $q$, and may
result in sharp transitions~---~the so-called \textit{pitch transitions}~--~between 
different branches of
metastable states.  The dependence of the twist wave number $q$ on
the free twisting number $q_0$ is then discontinuous. 

In particular, these discontinuities 
manifest themselves in
a jump-like temperature dependence of selective
light transmission 
spectra~\cite{Zink:jetp:1997,Gandhi:pre:1998,Zink:1999,Yoon:lc:2006}.
Different mechanisms behind the temperature variations of the pitch in CLC cells and hysteresis
phenomena
were discussed in Refs.~\cite{Bel:eng:2000,Bel:eng:2003,Palto:jetp:2002}.
A comprehensive stability analysis
of the helical structures  in CLC cells with
symmetric and asymmetric boundary conditions
was performed in Ref.~\cite{Kiselev:pre-1:2005}.
The effects of bistable surface anchoring and mechanical strain
on the pitch transitions
have been studied theoretically in the recent papers~\cite{McKay:epje:2012}
and~\cite{Lelidis:pre:2013}, respectively.

In practice, most cholesteric liquid crystals are
prepared on the basis of nematic LC mixtures
doped with chiral additives that induce a helical structure~\cite{Oswald:bk:2005}.
For photosensitive chiral dopants,
their helical twisting power and thus
the equilibrium helix pitch may, in principle, be controlled 
by light giving rise to the technologically promising effect of phototunable selective
reflection
(i.e. a change in the spectral position of the bandgap with
light exposure)~\cite{Vinograd:mclc:1990,Kurihara:apl:1998,
White:jap:2010,Kosa:nature:2012,Vernon:optexp:2013}.
The mechanism underlying the phototunable reflection
typically involves photoinduced changes in 
dopant conformation that affect the LC's helical twisting power
(see the recent review~\cite{Eelkema:lc:2011}). 

On the other hand,
the light-driven variations of the free twist wave number
may trigger the  pitch transitions discussed above
and can be used as a tool to explore
the details of such transitions,
depending on a variety of factors.
In particular, the surface anchoring energy
is known to have a profound effect
on the pitch transitions.
These surface mediated effects will be of our primary concern.

More specifically, we shall study the pitch transitions
in photosensitive CLC cells with strong
and semistrong anchoring conditions
by using an experimental method
that involves modeling of the photoreaction kinetics
combined with polarimetry measurements.
The results of modeling of the
photoreaction kinetics are used to obtain the
equilibrium pitch $P_0$ 
as a function of the uv irradiation time.
A similar irradiation time 
dependence of the pitch in the CLC cells, $P$, 
is extracted from experimental data on
the ellipticity of transmitted light measured
at different irradiation doses.
The resulting dependence of 
the twist wave number $q$
on the free twisting number $q_0$
describes the pitch transitions and can be interpreted
using known theoretical models.
 
The paper is organized as follows.
Experimental details are given in 
Sec.~\ref{sec:experiment}, where we describe
the materials and the methods of measurements.
In Sec.~\ref{sec:results}, we present the experimental 
data and apply the theoretical results~\cite{Kiselev:pre-1:2005} 
to interpret them.
Concluding remarks are given in Sec.~\ref{sec:disc-concl}.
Technical details on the method used to
compute the ellipticity of light transmitted through
CLC cells are relegated to Appendix~\ref{sec:norm-incid}.

\begin{figure*}[!tbh]
\centering
\resizebox{110mm}{!}{\includegraphics*{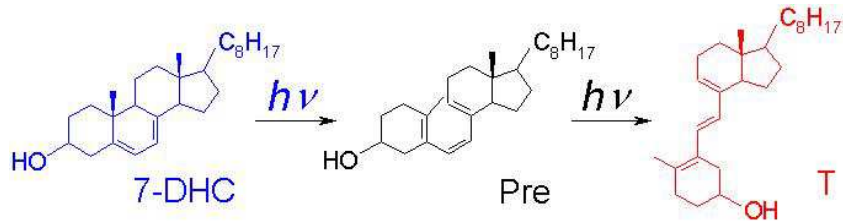}}
\caption{%
(Color online)
Scheme of key 7-DHC phototransformations in a nematic LC matrix under
uv irradiation at the wavelength $\lambda_{uv}= 254$~nm
(see Refs.~\cite{Orlova:optsp:2006,Orlova:optsp:2010} for more details).
7-DHC is provitamin D$_3$, 
Pre indicates previtamin D, and T stands for tachysterol.
}
\label{fig:chem}
\end{figure*}

\section{Experiment}
\label{sec:experiment}

In this section we describe 
the samples and the experimental
technique used to estimate both 
the equilibrium pitch, $P_0$,
and the pitch characterizing the helical structures 
formed in the CLC cells.
For this purpose,
in Sec.~\ref{subsec:samples}, 
the method
of uv absorption spectroscopy 
is used in combination with modeling the kinetics of 7-DHC
photoreaction to determine
the concentrations of photoisomers
that govern the equilibrium pitch.
In Sec.~\ref{subsec:polarimetry},
we present the experimental results 
for the ellipticity of  light transmitted through the cells
that are used to estimate
the pitch in the CLC cells.

\subsection{Photokinetics of equilibrium pitch}
\label{subsec:samples}

As a system with light controlled CLC pitch, we have 
used the nematic MLC-6815 (Merck) doped with 
the uv-sensitive right-handed chiral dopant provitamin $D_3$ 
(7-dehydrocholesterol, 7-DHC)
with the helical twisting power 
$HTP = +3.5$~\mum$^{-1}$wt.$^{-1}$.
Under the action of uv irradiation 
this dopant (provitamin $D_3$) is known to
undergo transformation into the left-handed \textit{trans}-isomer tachysterol with 
$HTP = -8.5$~\mum$^{-1}$wt.$^{-1}$~\cite{Orlova:mclc:2011}. 
By contrast, the nematic mixture MLC-6815 is uv-transparent
at wavelengths ranged from $240$~nm to $400$~nm
and the liquid crystal host remains stable under such uv irradiation.

The kinetics of the 7-DHC photoreaction is detailed
in Refs.~\cite{Jacobs:adph:1979,Saltiel:hndbook:2004}. 
It is well known that, in ethanol solution, the efficiency of 7-DHC conversion to
the \textit{trans}-isomer tachysterol  under uv irradiation at the
wavelength $\lambda_{uv}=254$~nm is about 60\%. This
photo-transformation is
thermally irreversible and is
accompanied by increase of the absorption maximum at the
wavelength $282$~nm. This increase
can be measured using the method of uv absorption spectroscopy and 
the results can be used for an indirect assessment of the tachysterol concentration.

In nematic LCs, the efficiency of tachysterol accumulation strongly
depends on the initial 7-DHC
concentration~\cite{Orlova:optsp:2006,Orlova:optsp:2010,Orlova:mclc:2011}.  
In our experiment,
the initial 7-DHC concentration was 
$C_{7-DHC}\approx 0.4$~wt.\%.
At this concentration,
we have 100\% efficiency of 7-DHC conversion to
tachysterol.
For this case,
the photochemical transformations of 7-DHC 
are schematically illustrated in Fig.~\ref{fig:chem}.
Note that we additionally controlled the
photoreaction efficiency by performing measurements of the uv absorption
spectra before irradiation and at the time corresponding to maximum
increase of absorption at the wavelength $282$~nm 
(it typically takes about 6 min).

\begin{figure*}[!tbh]
\centering
\resizebox{100mm}{!}{\includegraphics*{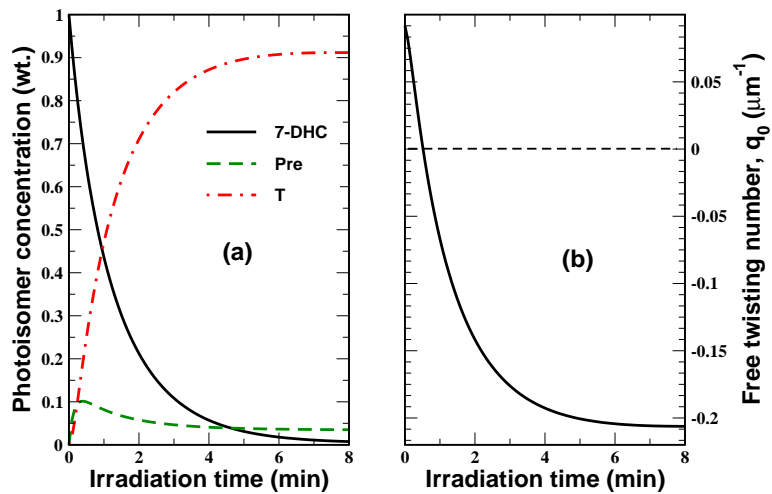}}
\caption{%
(Color online)
(a)~Photoisomer
concentrations and (b)~equilibrium cholesteric wave number
$q_0=2\pi/P_0$ computed as a function of the irradiation time.
}
\label{fig:photo_kin}
\end{figure*}

For the simplified scheme shown in Fig.~\ref{fig:chem},
the temporal evolution of the photoisomer concentrations
can be evaluated using the  kinetic model of the 7-DHC 
photoreaction developed in Ref.~\cite{Terenetskaya:jpbb:1999}.
The concentrations computed as a function of the uv
irradiation time are shown in Fig.~\ref{fig:photo_kin}(a).
The important point is that, in our calculations, 
the effect of the liquid crystal host 
on the quantum yields of phototransformations
is taken into account.

According to the well-known formula  
\begin{align}
  \label{eq:P0-concentr}
  P_0^{-1}=\sum_i w_i\,HTP_i\,  C_i,
\end{align}
where $w_i$ is the weight fraction of the $i$th chiral photoisomer,
and the equilibrium cholesteric pitch $P_0$  is 
determined by the photoisomer concentrations. 
The calculated concentrations can now be
substituted into Eq.~\eqref{eq:P0-concentr} 
to obtain the irradiation time dependence
of the free twisting wave number
depicted in Fig.~\ref{fig:photo_kin}(b).

In our experiments,
we have  used
the planar CLC cells of the thickness $D$
varied between $55$~\mum\ and
$65$~\mum.
At the initial 7-DHC concentration 
$C_{7-DHC}\approx 0.4$~wt.\%,
the photoinduced reorientation processes
in such cells
are not complicated by inhomogeneity
effects related to the formation of
highly twisted states.

\begin{figure*}[!hbt]
\centering
  \resizebox{70mm}{!}{\includegraphics*{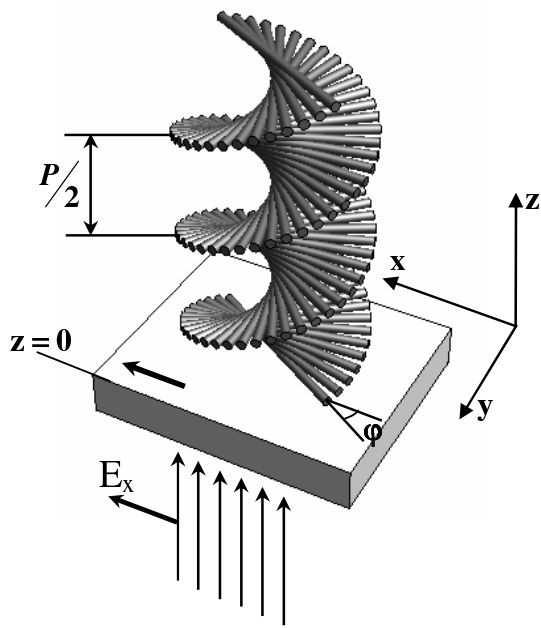}}
\caption{%
Geometry of normal incidence:
A plane wave of linearly polarized light is impinges on the CLC cell. 
}
\label{fig:norm_inc_geom}
\end{figure*}

\subsection{Polarimetry measurements}
\label{subsec:polarimetry}

When a light beam
propagates through
an optically anisotropic medium,
the anisotropy is known to greatly affect
its state of polarization~\cite{Born:bk:1999}.
This state is generally described by 
the Stokes parameters
and can be conveniently represented by 
a polarization ellipse
whose orientation and eccentricity are specified by the azimuthal angle
of polarization (\textit{polarization azimuth}) $\phi_p$ and 
the \textit{ellipticity} $\epsilon_{\ind{ell}}$, 
respectively~\cite{Azz:1977,Goldstein:bk:2003,Tomp:bk:2005}. 

For  light propagating through a CLC cell where
the optical anisotropy is determined by the
helical orientational structure~\eqref{eq:director},
its ellipticity is sensitive to 
the pitch of 
the CLC spiral~\cite{Belyakov:bk:1989}.
Thus the cholesteric pitch $P$ in photosensitive CLC cells 
may, in principle, be estimated by measuring
the ellipticity of light passed through the cells.

In our experiments, the measurements were performed
for light which is normally incident onto the aligned substrate
and is linearly polarized along the direction of rubbing. 
Figure~\ref{fig:norm_inc_geom} illustrates the geometry of normal incidence. 
 
We have used
planar CLC cells
where the photosensitive CLC was sandwiched 
between quartz substrates.
In the symmetric case of strong anchoring conditions, both
the quartz substrates were coated with rubbed 
polydimethylsiloxane
aligning layers which are insensitive to the uv irradiation. 
We have also examined asymmetric CLC cells with
semistrong anchoring conditions.
These cells were assembled using 
the exiting substrate without the aligning coating.

\begin{figure*}[!hbt]
\centering
   \resizebox{100mm}{!}{\includegraphics*{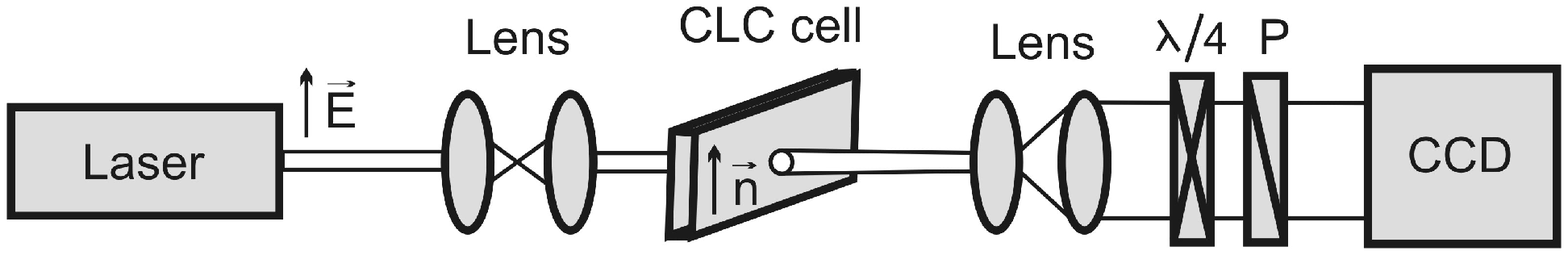}}
\caption{%
Scheme of the polarimeter. Setup consists of a He-Ne laser
($\lambda=633$~nm), collimating lenses, 
CLC cell, beam-expander, Stokes analyzer (quarter-wave plate and
polarizer) and charge coupled device (CCD) camera.
}
\label{fig:setup}
\end{figure*}

\begin{figure*}[!tbh]
\centering
\resizebox{140mm}{!}{\includegraphics*{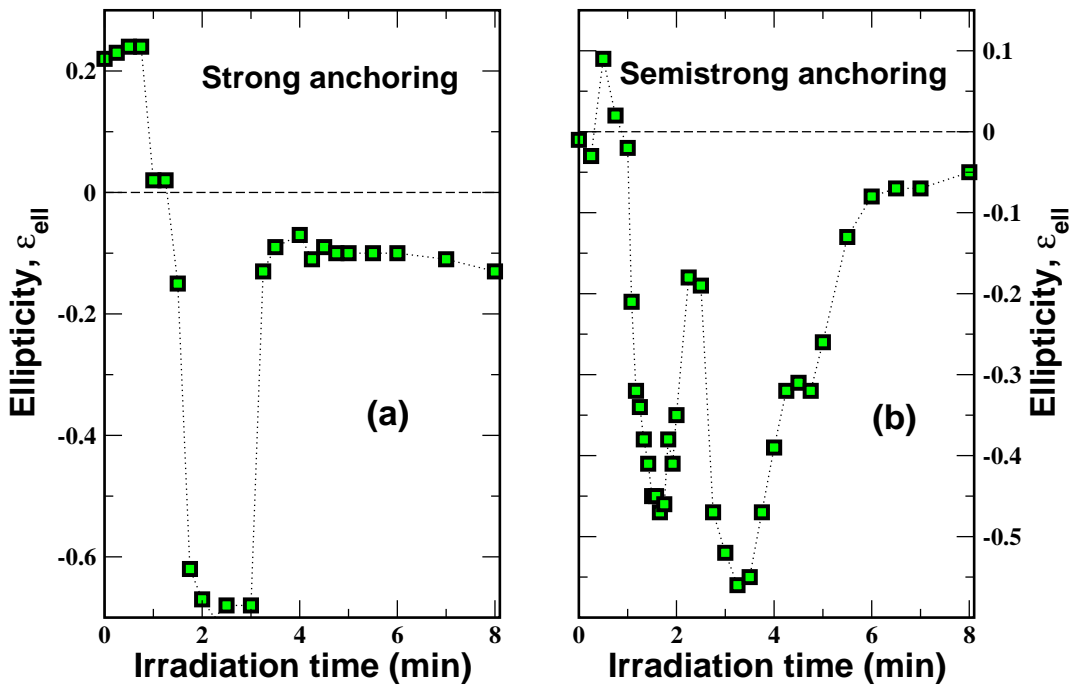}}
\caption{%
(Color online)
(a)~
Ellipticity $\epsilon_{\ind{ell}}$
of transmitted light measured as a function
of uv irradiation time
in cells with
(a)~strong and (b)~semistrong
anchoring conditions. 
}
\label{fig:ellipt-vs-t}
\end{figure*}

After each step of uv irradiation,
the ellipticity of
light transmitted through the cell 
was measured using the standard
Stokes polarimetry technique
which is described in our previous 
papers~\cite{Kiselev:pra:2008,Kiselev:sid:2006}.
The time interval between irradiation and
polarimetry studies was long enough 
(up to 30 min)
to allow for the processes of reorientation
to reach the stationary state. 

Figure~\ref{fig:setup} shows
the setup scheme used in our experiments.
Referring to Fig.~\ref{fig:setup}, 
the cell is irradiated with a beam
generated by a He-Ne laser
(the wavelength is $633$~nm)
and passed through the collimating lenses.
After the cell,  the beam is expanded
and a charge coupled device (CCD) camera
collects
the output from the Stokes analyzer
represented by the combination of the quarter wave plate
and the polarizer.

Figure~\ref{fig:ellipt-vs-t} presents 
the results for the ellipticity measured 
at different irradiation doses in the
symmetric and asymmetric CLC cells. 
These results were derived
using the standard
procedure~\cite{Born:bk:1999,Azz:1977,Goldstein:bk:2003,Tomp:bk:2005}
which involves
performing the intensity measurements
at six different combinations of the quarter
wave plate and the polarizer
needed to obtain the Stokes parameters.


\begin{figure*}[!tbh]
\centering
\resizebox{140mm}{!}{\includegraphics*{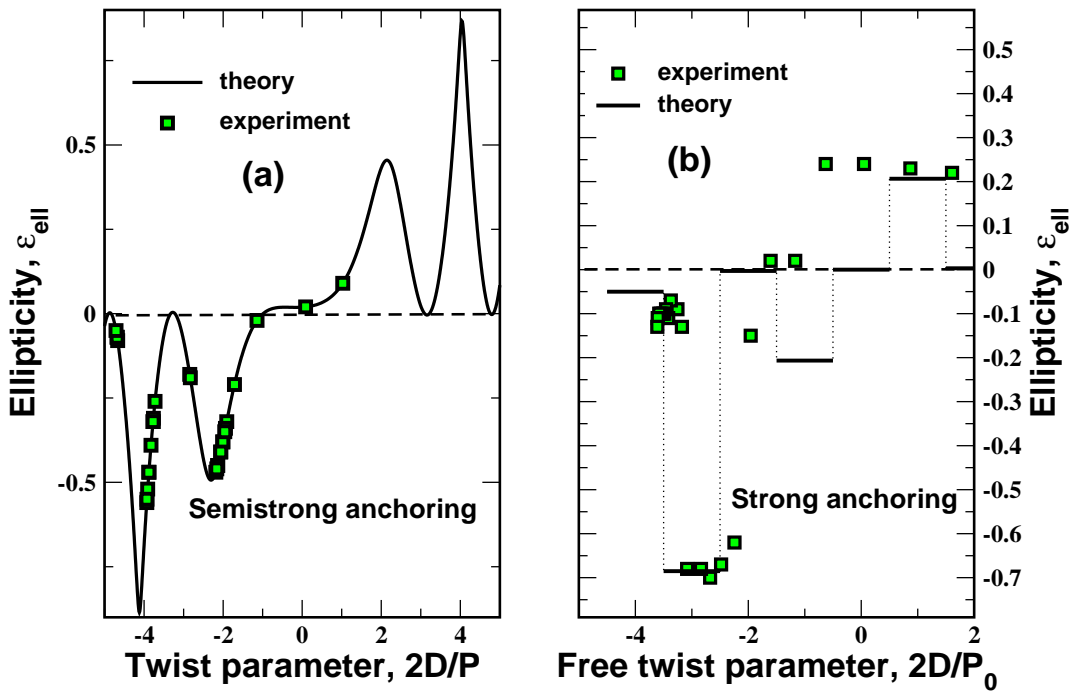}}
\caption{%
(Color online)
(a)~Ellipticity $\epsilon_{\ind{ell}}$
computed as a function of the twist parameter,
$2D/P$,
for light transmitted through a CLC cell
of thickness $D = 62$~\mum.
$\Delta\phi=3$~deg (see Eq.~\eqref{eq:transm_wave})
and
$n_{\perp}=1.4674$ ($n_{\parallel}=1.5191$)  is the ordinary (extraordinary)
refractive index.
Squares indicate the places that are associated with the experimental points
by applying the procedure described in Sec.~\ref{subsec:asymmetric-cells}.
(b)~Ellipticity $\epsilon_{\ind{ell}}$
computed for seven equilibrium helical structures
($2D/P\in\{-4,-3,-2,-1,0,1,2\}$) in the symmetric CLC cell
for $\Delta\phi=0$~deg.
Squares represent the $q_0$-dependence
of the ellipticity obtained for 
the experimental data
shown in Figs.~\ref{fig:ellipt-vs-t}(a)
and~\ref{fig:photo_kin}(b).
}
\label{fig:ellipt_azim}
\end{figure*}

The theoretical results shown in Fig.~\ref{fig:ellipt_azim}
are computed from the analytical expression for 
the transmission matrix deduced in Appendix~\ref{sec:norm-incid}
(see formulas~\eqref{eq:T_norm_inc},~\eqref{eq:TR_rw}
and~\eqref{eq:W_rw_11})
using the transfer matrix method
in the form formulated in Refs.~\cite{Kiselev:jpcm:2007,Kiselev:pra:2008}.
In particular, the curve depicted 
in Fig.~\ref{fig:ellipt_azim}(a)
represents the $q$-dependence of the
ellipticity
and can be used to estimate the helix pitch $P$
at different irradiation doses by making comparison
between the experimental data and the results of calculations.  
In the subsequent section we provide 
details on the procedure used for data processing
and discuss the results.

\section{Results}
\label{sec:results}

\begin{figure*}[!tbh]
\centering
  \resizebox{120mm}{!}{\includegraphics*{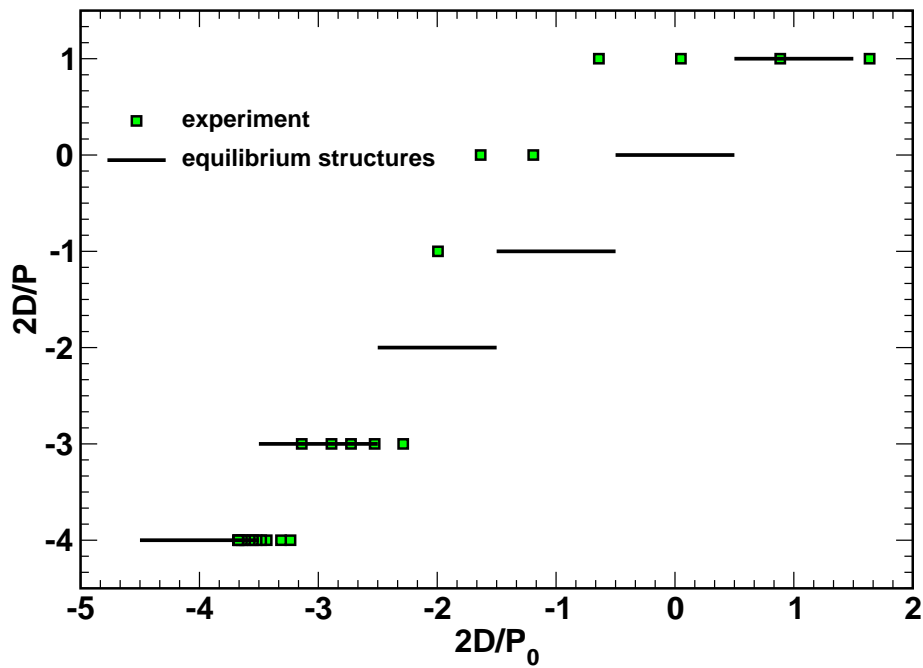}}
\caption{%
(Color online)
Twist parameter $2D/P$
versus free twist parameter  $2D/P_0$
measured in the CLC cell with strong anchoring conditions.
}
\label{fig:strong}
\end{figure*}

At this stage, our task is to evaluate the pitch
of the CLC helical structure formed at different irradiation doses
from the results of the previous section.
In this section, we detail the procedures used for this purpose
and present the results.

\subsection{Strong anchoring: Symmetric cells}
\label{subsec:symmetric-cells}

When the anchoring is strong at both substrates,
the boundary conditions require the CLC director~\eqref{eq:director} 
at the substrates to be parallel to the corresponding easy axis
(in experiments, the easy axes are defined by the direction of rubbing).
Owing to the boundary conditions,
the helix wave number $q$ takes values from a discrete set.
This set represents the helical structures characterized
by the twist parameter $\nu=q D/\pi=2 D/P$
and labeled by the half-turn number $k$,
\begin{align}
  \label{eq:strong}
  \nu\equiv qD/\pi=k,
\quad
k\in \mathbb{Z}.
\end{align}
 The equilibrium value of $k$
is the integer that minimizes
the distance between $k$
and the free twist parameter $\nu_0=q_0 D/\pi=2 D/P_o$.
The resulting step-like dependence
of $2 D/P$ on $2 D/P_0$
for the equilibrium helical structures 
is depicted in Fig.~\ref{fig:strong}.

According to the stability analysis
of Ref.~\cite{Kiselev:pre-1:2005},
instability caused by slippage of
the director in the plane of the spiral
cannot occur provided the azimuthal anchoring
is strong at both substrates.
The structures may, however,
lose their stability due to
out-of-plane fluctuations.
It was shown that, when the energy cost of bending is relatively
small, the structure becomes unstable
at sufficiently large distance between
its wave number $q$ and $q_0$~\cite{Kiselev:pre-1:2005}.

The values of the ellipticity
indicated in Fig.~\ref{fig:ellipt_azim}(b)
are calculated for 
helical structures in
the symmetric cell
with strong anchoring conditions.
From Eq.~\eqref{eq:strong},
these structures are characterized
by the integer half-turn number $2 D/P=k$
which is independent of the free twisting wave number.

The experimental points in Fig.~\ref{fig:ellipt_azim}(b)
represent 
the dependence of the ellipticity
on the free twist parameter $2 D/P_0$
that can be obtained 
from the data shown in Fig.~\ref{fig:ellipt-vs-t}(b)
with the help of
the irradiation time dependence of 
the free twisting wave number
shown in Fig.~\ref{fig:photo_kin}(b).
These points can now be related to
the half-turn number $k$,
by minimizing both 
the difference between the theoretical and experimental values
of the ellipticity
and the change in
the half-turn number $\Delta k$. 

Figure~\ref{fig:strong} shows the experimental 
$q_0$-dependence of the helix twist parameter 
$2 D/P$
measured in the CLC cells with strong anchoring conditions
at both substrates.
Referring to Fig.~\ref{fig:strong},
it can be seen that the experimental data
indicate the presence of metastable states
and jump-like transitions with $\Delta k =1$ and $\Delta k =2$.

\subsection{Semistrong anchoring: Asymmetric cells}
\label{subsec:asymmetric-cells}

Asymmetric CLC cells represent
the case of mixed boundary conditions
in which the strong anchoring limit applies
only to the entrance plate, $z=0$.
This case is referred to as semistrong anchoring
and we assume that
the anchoring potential at the substrate 
with weak anchoring conditions
can be taken in
the Rapini-Papoular form~\cite{Rap:1969}:
\begin{align}
  \label{eq:R_P}
  V_s(\phi_s)=\frac{W}{2}\sin^2(\phi_s-\phi_e),
\end{align}
where $W$ is the anchoring energy strength,
$\phi_s\equiv \phi(D)$ is the director azimuthal angle at the surface
and $\phi_e$ is the azimuthal angle of the easy axis,
$\uvc{e}=(\cos\phi_e,\sin\phi_e,0)$.

For such CLC cells,
the relation between 
the helix wave number
and the free twisting wave number
can be conveniently written in the following form~\cite{Kiselev:pre-1:2005}:
\begin{align}
&
  \label{eq:beta_0-vs-beta}
  \nu_0=\nu + w/\pi \sin 2(\pi\nu-\phi_e),
\quad
w=\frac{W D}{2 K_t},
\notag
\\
&
\nu=2 D/P,
\quad
\nu_0=2 D/P_0,
\end{align}
where $K_t$ is the twist elastic constant.
The stability condition for the helical configurations characterized
by the twisting parameter $\nu$
is given by
\begin{align}
  \label{eq:stability}
  1 + 2 w \cos 2(\pi\nu-\phi_e)>0.
\end{align}

\begin{figure*}[!tbh]
\centering
  \resizebox{140mm}{!}{\includegraphics*{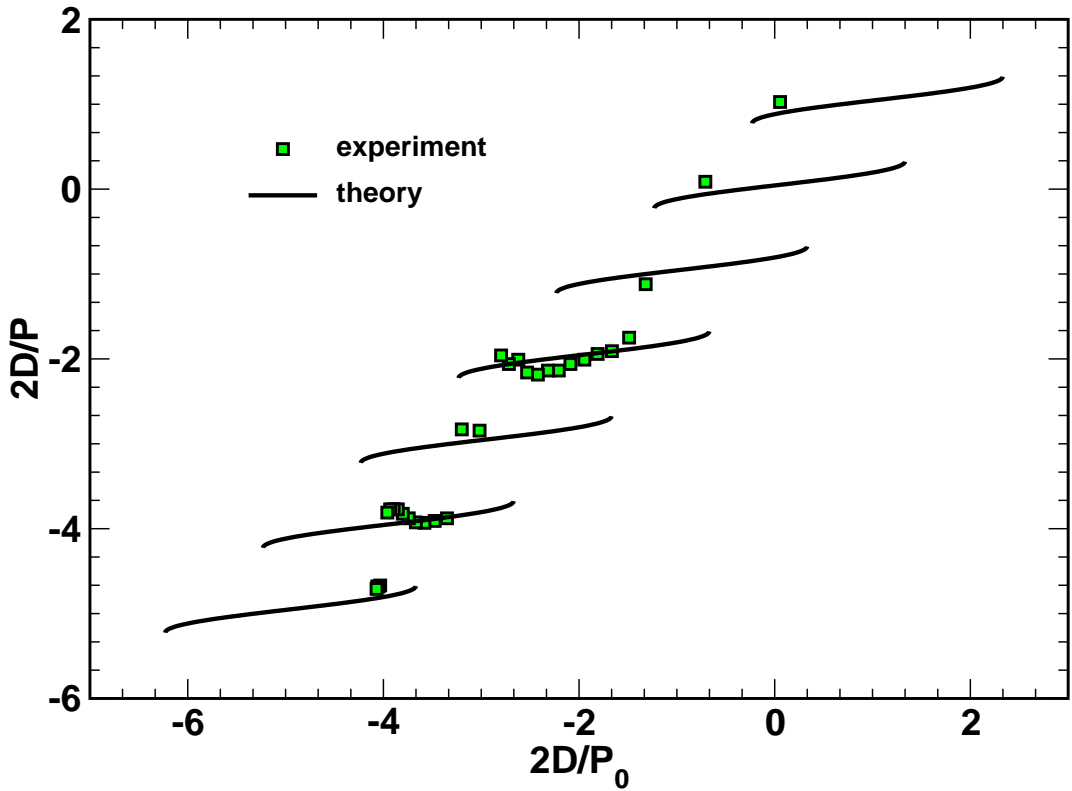}}
\caption{%
(Color online)
Dependence of the twist parameter, $2D/P$,
on the free twist parameter,  $2D/P_0$,
measured in the CLC cell with semistrong anchoring conditions.
Theoretical curve is computed from Eq.~\eqref{eq:beta_0-vs-beta}
at $w=3.2$ and $\phi_e=9$~deg.
Solid line represents branches of stable twisting states
that meet the stability condition~\eqref{eq:stability}.
}
\label{fig:semistrong}
\end{figure*}

Formulas~\eqref{eq:beta_0-vs-beta} and~\eqref{eq:stability}
can be used for processing the experimental data
presented in Fig.~\ref{fig:ellipt-vs-t}(b). 
This procedure produces
dependence of the twist parameter $2D/P$
on the free twisting parameter $2 D/P_0$
based on the data shown in Figs.~\ref{fig:ellipt-vs-t}(b) and~\ref{fig:photo_kin}(b)
and the ellipticity computed as a function of $2D/P$ 
[see the theoretical curve in Fig.~\ref{fig:ellipt_azim}(a)].
It works as follows:
\renewcommand{\theenumi}{\alph{enumi}}
\renewcommand{\labelenumi}{(\theenumi)}
\begin{enumerate}
\item 
For each point in Fig.~\ref{fig:ellipt-vs-t}(b),
the value of the irradiation time is used to compute
the corresponding value of the free twisting wave number
using the curve depicted in Figs.~\ref{fig:photo_kin}(b).

\item 
For each value of the measured ellipticity in
Fig.~\ref{fig:ellipt-vs-t}(b)
and the associated free twisting parameter,
we generally obtain multiple values of the twist parameter
representing the points 
on the theoretical curve in Fig.~\ref{fig:ellipt_azim}(a)
with ellipticity equal to the measured one.
The next step describes the selection procedure.

\item 
Given the free twisting parameter
and the values of the twist parameter, $2 D/P_0=\nu_0$ 
and $2 D/P=\nu$, 
we evaluate $1 + 2 w \cos 2(\pi\nu-\phi_e)$ and
the difference $\Delta=|\nu_0-\nu -w/\pi \sin 2(\pi\nu-\phi_e)|$.
Then we choose the twist parameter that 
satisfies the stability condition~\eqref{eq:stability}
and minimizes $\Delta$.
The selected points are indicated by 
squares in Fig.~\ref{fig:ellipt_azim}(a). 

\item 
The result is that each point in Fig.~\ref{fig:ellipt-vs-t}(b) 
is characterized by the free twisting parameter
and the twist parameter.
These parameters define the points 
indicated by squares in Fig.~\ref{fig:semistrong}.
\end{enumerate}
  
From Fig.~\ref{fig:semistrong},
it can be seen that
the results for the asymmetric cell 
are in good agreement with the theoretical
predictions of Ref.~\cite{Kiselev:pre-1:2005}
(Eqs.~\eqref{eq:beta_0-vs-beta} and ~\eqref{eq:stability}).
They indicate that the jump-like pitch transitions
occur only between the adjacent branches of
stable helical structures,
unlike in the case of symmetric cell with strong anchoring.

Note that the above procedure
relies on the computed
curve representing the $q$-dependence of
the ellipticity of light transmitted through
the CLC cell.
In addition to parameters such as 
the cell thickness and the refractive indices,
this curve depends on the angle between 
the surface director at the entrance substrate 
and the polarization vector of the linearly polarized incident light, $\Delta\phi$.
It is found that the best fit value of this angle is about $3$~deg.
This implies that, in the asymmetric cells, 
the surface director may not be parallel to the
rubbing direction. A similar phenomenon was previously reported in 
Ref.~\cite{Yoon:lc:2006}.
In our case, however, the angle is relatively small
and can be attributed to the misalignment error.

In closing this section, we briefly comment
on the estimated value of the anchoring energy parameter,
$w\approx 3.2$.
For the thickness $D\approx 62$~\mum\
and the twist elastic constant  $K_t\approx 10^{-12}$~N,
the anchoring energy strength can be estimated
at about $W= 2 w K_t/D\approx 10^{-7}$~J/m$^2$.
It comes as no surprise that, for the untreated substrate, 
the estimated anchoring energy
is at least two orders of magnitude smaller than typical values for the azimuthal
anchoring energy strength~\cite{Kiselev:pre-2:2005}.

\section{Discussion and conclusions}
\label{sec:disc-concl}

In the confined geometry of planar cells, 
the helical structures formed in the cells
and their stability
are greatly affected by
the boundary conditions imposed
at the confining surfaces.
The helix pitch characterizing these structures
generally differs from its equilibrium value.
A more important additional effect
is the presence of
multiple metastable twisting states in such cells, 
which appear
as a result of interplay between the bulk and the surface
contributions to the free energy.  
Changes of the equilibrium pitch
may trigger sharp transitions~---~the so-called
pitch transitions~---~between 
different branches of metastable states.

In this paper we have studied the pitch transitions
in cells filled with photosensitive chiral nematic liquid crystals.
In such materials, the equilibrium pitch can be 
efficiently controlled by light through photochemically 
induced transformations of chiral dopants.

In order to determine
the concentrations of photoisomers
that govern the equilibrium pitch
(and the free twisting wave number $q_0=2\pi/P_0$)
we have used
the method
of uv absorption spectroscopy 
combined with modeling 
the kinetics of the photoreaction.
In our experiments, the free twisting
wave number  is found to be a monotonically
decreasing function of irradiation time
[see Fig.~\ref{fig:photo_kin}(b)].

The pitch of helical structures formed in the cells
after each step of irradiation
was estimated from the experimental results 
of polarimetry measurements
giving the ellipticity of light transmitted through
the cells at different irradiation doses (see Fig.~\ref{fig:ellipt-vs-t}).
There are two cases that have been studied experimentally:
(a)~a symmetric cell with strong anchoring conditions
at both substrates; and (b)~an asymmetric cell with mixed
boundary conditions where weak anchoring conditions
are applied at one of the substrates (semistrong anchoring).

From the 
steplike dependence of the twist parameter
$\nu=2D/P$ on the free twist parameter
$\nu_0=2D/P_0$
shown in Fig.~\ref{fig:strong},
it can be concluded that
the light-induced pitch transitions in the symmetric cell
are governed by the boundary conditions
and involve metastable twisting states.
By contrast, the similar dependence for 
the asymmetric cell with semistrong anchoring
(see Fig.~\ref{fig:semistrong})
shows successive jumplike transitions
that take place
between the branches of stable twisting states
where the twist parameter $\nu$
monotonically increases with $\nu_0$.

We have found that such behavior agrees very well with
the predictions of the theoretical analysis 
performed in Ref.~\cite{Kiselev:pre-1:2005}.
According to this analysis,
the helical structure responds to variations of the
free wave number $q_0$ (and thus the free twist parameter $\nu_0$) 
by changing its twist parameter $\nu$.
This change may render the initially equilibrium structure
either metastable or unstable.
Under certain conditions, this instability
is governed by in-plane director fluctuations.
The mechanism dominating
transformations of the director field then can be described as
director slippage through the energy barriers
formed by the surface potentials.

For the case of semistrong anchoring,
equations~\eqref{eq:beta_0-vs-beta}
and~\eqref{eq:stability} 
define branches of stable helical structures.
These formulas were used to fit the experimental data
and 
the best fit value of the anchoring energy parameter
is estimated at about $w= WD/(2K_t)\approx 3.2$.
So, the anchoring energy strength at the untreated substrate
is found to be at least two orders of magnitude smaller than typical values
of the azimuthal anchoring strength.
It turns out that this value is not small enough
to suppress the jump-like behavior.
From Eq.~\eqref{eq:stability}, 
the latter occurs at $w<1$.

In conclusion, it should be emphasized that
the non-equilibrium dynamics of 
the light-induced pitch transitions
is well beyond the scope of this paper.
We have demonstrated that 
use of
photosensitive CLCs
with light controlled equilibrium pitch
provides a useful tool for investigation of
such transitions and 
we hope that our study will stimulate further
progress in this field. 


\appendix

\section{Optics of helical structures at normal incidence: exact solution revisited}
\label{sec:norm-incid}

In this appendix we
briefly outline the transfer matrix approach 
in the form formulated
in Refs.~\cite{Kiselev:jpcm:2007,Kiselev:pra:2008,Kiselev:pre:2011} 
and show how it can be used to describe 
the optical properties 
of ideal CLC helical structures.
The director field of these structures
is given in Eq.~\eqref{eq:director}
and is  
characterized by the helix wave number
$q=2 \pi/P$, where $P$ is the CLC pitch.

\subsection{Transfer matrix method}
\label{subsec:transf-matrix}

We deal with a harmonic electromagnetic field 
characterized by the free-space wave number
$k_{\vac}=\omega/c$,
where $\omega$ is the frequency
(the time-dependent factor
is $\exp\{-\omega t\}$),
and consider the slab geometry. 
In this geometry, an optically anisotropic layer
of thickness $D$
is sandwiched between 
the bounding surfaces (substrates): $z=0$ and $z=D$
(the $z$ axis is normal to the substrates)
and is
characterized by the dielectric tensor $\epsilon_{ij}$
and the magnetic permittivity $\mu$.
The dielectric tensor can be expressed
in terms of the director~\eqref{eq:director}
as follows
\begin{align}
  \label{eq:diel_clc}
  \epsilon_{ij}(z)=\epsilon_{\perp}\delta_{ij}+
\Delta\epsilon\, n_{i}(z)\, n_{j}(z), 
\quad
\Delta\epsilon=
\epsilon_{\parallel}-
\epsilon_{\perp},
\end{align} 
where $\delta_{ij}$ is the Kronecker symbol
and $n_{\perp}=\sqrt{\mu \epsilon_{\perp}}$
($n_{\parallel}=\sqrt{\mu \epsilon_{\parallel}}$)
is the ordinary (extraordinary) refractive index.

Further, 
we restrict ourselves to the case of stratified media
and assume that
the electromagnetic fields can be taken in the following
factorized form
\begin{align}
  \label{eq:EH-form}
\{\vc{E}(\vc{r}), \vc{H}(\vc{r})\}=
\{\vc{E}(z), \vc{H}(z)\}\exp\sca{\vc{k}_p}{\vc{r}},
\end{align}
where the vector
\begin{align}
  \label{eq:k_p}
  \vc{k}_p/k_{\vac}=\vc{q}_p=
q_p(\cos\phi_p,\sin\phi_p,0)
\end{align}
represents the lateral component of the wave vector.
Then we write down
the representation for the electric and magnetic fields, $\vc{E}$ and
$\vc{H}$,
\begin{align}
  \label{eq:decomp-E}
  \vc{E}=E_z \uvc{z} +\vc{E}_{P},\quad
\vc{H}= H_z \uvc{z} +\uvc{z}\times \vc{H}_{P},
\end{align}
where the  components directed along the normal to the bounding surface
(the $z$ axis) are separated from the tangential (lateral) ones. 
In this representation,
the vectors
$\vc{E}_{P}=E_x \uvc{x}+E_y \uvc{y}\equiv
\begin{pmatrix}
  E_x\\E_y
\end{pmatrix}
$
and
$\vc{H}_{P}=\vc{H}\times\uvc{z}\equiv
\begin{pmatrix}
  H_y\\-H_x
\end{pmatrix}
$
are parallel to the substrates
and give the lateral components of the electromagnetic field.

Substituting the relations~\eqref{eq:decomp-E}
into the Maxwell equations and
eliminating the $z$ components of the electric and 
magnetic fields gives
equations for 
the tangential components of the electromagnetic field
that can be written  
in the following $4\times 4$ matrix 
form~\cite{Kiselev:pra:2008}:
\begin{align}
  \label{eq:matrix-system}
  -i\pdrs{\tau}\vc{F}=\mvc{M}\,\vc{F}\equiv
    \begin{pmatrix}\mvc{M}_{11}&\mvc{M}_{12}\\\mvc{M}_{21}&\mvc{M}_{22} \end{pmatrix}
    \begin{pmatrix}\vc{E}_{P}\\\vc{H}_{P} \end{pmatrix},
\:
\tau\equiv k_{\vac} z.
\end{align}
For the dielectric tensor~\eqref{eq:diel_clc}
with the plane of incidence parallel to the $x$-$z$ plane,
from the general expressions derived in 
Refs.~\cite{Kiselev:jpcm:2007,Kiselev:pra:2008},
the $2\times 2$ matrices $\mvc{M}_{ij}$ characterizing 
the block structure of the matrix $\mvc{M}$ are given by
\begin{align}
  \label{eq:matrix-M12}
  &
\mvc{M}_{12}=\mu\mvc{I}_2-\frac{q_p^2}{2\epsilon_{\perp}}
(\mvc{I}_2+\bs{\sigma}_3),
\quad
\mvc{M}_{ii}=\mvc{0},
\\
&
  \label{eq:matrix-M21}
 \mvc{M}_{21}=
-\frac{q_p^2}{2\mu}
(\mvc{I}_2-\bs{\sigma}_3)
+
\notag
\\
&
\epsilon_c
\Bigl\{
\mvc{I}_2+ 
u_a \bigl[
\cos(2\phi)\,\bs{\sigma}_3+
\sin(2\phi)\,\bs{\sigma}_1
\bigr]
\Bigr\},
\\
 \label{eq:epsl_c}
&
  \epsilon_c=(\epsilon_{\parallel}+\epsilon_{\perp})/2,
\quad
u_a=\frac{\epsilon_{\parallel}-\epsilon_{\perp}}{\epsilon_{\parallel}+\epsilon_{\perp}},
\\
&
\label{eq:phi}
\phi= \tilde{q} \tau +\phi_0,
\quad
\tilde{q}=q/k_{\vac}=\lambda/P,
\end{align}
where $\mvc{I}_n$ is the $n\times n$ identity matrix and
$\{\bs{\sigma}_1,\bs{\sigma}_2,\bs{\sigma}_3\}$ are the Pauli matrices given by
\begin{align}
  \label{eq:pauli}
      \bs{\sigma}_1=
      \begin{pmatrix}
        0&1\\1&0
      \end{pmatrix},
\:
      \bs{\sigma}_2=
      \begin{pmatrix}
        0&-i\\i&0
      \end{pmatrix},
\:
      \bs{\sigma}_3=
      \begin{pmatrix}
        1&0\\0&-1
      \end{pmatrix}.
\end{align}
General solution of the system~\eqref{eq:matrix-system}
\begin{align}
  \label{eq:evol-oprt}
  \vc{F}(\tau)=
  \mvc{U}(\tau,\tau_0)\,\vc{F}(\tau_0)
\end{align}
can be conveniently expressed in terms of
the \textit{evolution operator} defined as the matrix solution of 
the initial value problem
\begin{subequations}
  \label{eq:evol_problem}
\begin{align}
  \label{eq:evol_eq}
     -i\pdrs{\tau}\mvc{U}(\tau,\tau_0)
&
=
\mvc{M}(\tau)\,\mvc{U}(\tau,\tau_0),
\\
  \label{eq:evol_ic}
\mvc{U}(\tau_0,\tau_0)
&
=\mvc{I}_4,
\end{align}
\end{subequations} 

In the ambient medium with $\epsilon_{ij}=\epsilon_{\med}\delta_{ij}$
and $\mu=\mu_{\med}$, the general solution~\eqref{eq:evol-oprt}
can be expressed in terms of
plane waves propagating along  
the wave vectors with the tangential component~\eqref{eq:k_p}.
For such waves, the result  is given by~\cite{Kiselev:pre:2011}
\begin{align}
&
  \label{eq:F_med}
  \vc{F}_{\med}(\tau)=\mvc{V}_{\med}(\vc{q}_p)
  \begin{pmatrix}
    \exp\{i \mvc{Q}_{\med}\, \tau\} & \mvc{0}\\
\mvc{0} &    \exp\{-i \mvc{Q}_{\med}\, \tau\}   
  \end{pmatrix}
  \begin{pmatrix}
    \vc{E}_{+}\\
\vc{E}_{-}
  \end{pmatrix},
\\
&
\label{eq:Q_med}
\mvc{Q}_{\med}=q_{\med}\,\mvc{I}_2,
\quad
q_{\med}=\sqrt{n_{\med}^2-q_p^2},
\end{align}
where 
$\mvc{V}_{\med}(\vc{q}_p)$
is the eigenvector matrix for the ambient medium
given by
\begin{align}
&
\label{eq:Vm-phi-q}
\mvc{V}_{\med}(\vc{q}_p)=
\mvc{T}_{\ind{rot}}(\phi_p)\mvc{V}_{\med}=
\notag
\\
&
\begin{pmatrix}
 \mvc{Rt}(\phi_p)&\mvc{0}\\
\mvc{0}& \mvc{Rt}(\phi_p) 
\end{pmatrix}
\begin{pmatrix}
\mvc{E}_{\med} & -\bs{\sigma}_3 \mvc{E}_{\med}\\
\mvc{H}_{\med} & \bs{\sigma}_3 \mvc{H}_{\med}\\
\end{pmatrix},
\\
&
  \label{eq:EH-med}
  \mvc{E}_{\med}=
  \begin{pmatrix}
    q_{\med}/n_{\med}& 0\\
0 & 1
  \end{pmatrix},
\quad
 \mu_{\med}\,\mvc{H}_{\med}=
 \begin{pmatrix}
   n_{\med}& 0\\
0 & q_{\med}
 \end{pmatrix},
\\
&
\label{eq:Rot_matrix}
\mvc{Rt}(\phi)=\begin{pmatrix}
  \cos\phi &-\sin\phi\\
\sin\phi & \cos\phi
\end{pmatrix},
\end{align}

From Eq.~\eqref{eq:F_med},
the vector  amplitudes $\vc{E}_{+}$ and
$\vc{E}_{-}$ correspond to the forward and backward eigenwaves
with
$\vc{k}_{+}=k_{\vac}(q_{\med}\,\uvc{z}+\vc{q}_p)$
and 
$\vc{k}_{-}=k_{\vac}(-q_{\med}\,\uvc{z}+\vc{q}_p)$, respectively.
In the half space $z\le 0$ 
before the entrance face of the layer $z=0$,
these eigenwaves describe 
the \textit{incident and reflected waves}
 \begin{align}
   &
   \label{eq:in-out-plus}
   \vc{E}_{+}\vert_{z\le 0}=
\vc{E}_{\ind{inc}},
\quad
   \vc{E}_{-}\vert_{z\le 0}=
   \vc{E}_{\ind{refl}},
 \end{align} 
whereas,
in the half space $z\ge D$ after the exit face of the layer,
these waves are given by
\begin{align}
   \label{eq:in-out-minus}
   \vc{E}_{+}\vert_{z\ge D}=
\vc{E}_{\ind{trm}},
\quad
   \vc{E}_{-}\vert_{z\ge D}= 0,
\end{align}
where $\vc{E}_{\ind{trm}}$ is the vector amplitude of
the \textit{transmitted wave}.
The standard linear input-output relations
\begin{align}
&
  \label{eq:transm-rel}
\vc{E}_{\transm}
=\mvc{T}\,
\vc{E}_{\inc},
\quad
\vc{E}_{\refl}
=
\mvc{R}\,
\vc{E}_{\inc}
\end{align}
link the vector amplitudes of transmitted and reflected waves
$\vc{E}_{\transm}$ and $\vc{E}_{\refl}$ 
with the amplitude of the incident wave $\vc{E}_{\inc}$
through the transmission and reflection
matrices $\mvc{T}$ and $\mvc{R}$.

It is our task now to relate
these matrices and the evolution operator
given by Eq.~\eqref{eq:evol_problem}.
To this end, we 
use  the boundary conditions requiring
the tangential components of the electric and magnetic
fields to be continuous at the boundary surfaces:
$\vc{F}(0)=\vc{F}_{\med}(0-0)$ and
$\vc{F}(h)=\vc{F}_{\med}(h+0)$,
and apply the relation~\eqref{eq:evol_problem}
to  the anisotropic layer of the thickness $D$
to yield the following result
\begin{align}
  \label{eq:continuity}
  \vc{F}_{\med}(h+0)=\mvc{U}(h,0)\,\vc{F}_{\med}(0-0),
\quad  
h=k_{\vac} D.
\end{align}

On substituting Eqs.~\eqref{eq:F_med}
into Eq.~\eqref{eq:continuity}
we have
\begin{align}
  \label{eq:transf-rel}
  \begin{pmatrix}
    \vc{E}_{\ind{inc}}\\
    \vc{E}_{\ind{refl}}
  \end{pmatrix}
=
\mvc{W}\,
  \begin{pmatrix}
    \vc{E}_{\ind{trm}}\\
    \vc{0}
  \end{pmatrix}
\end{align}
where the matrix $\mvc{W}$
linking the electric field vector amplitudes of the waves
in the half spaces $z<0$ and $z>D$
bounded by the faces of the layer
will be referred to as the \textit{transfer (linking) matrix}.
The expression for the transfer matrix is
as follows
\begin{align}
&
  \label{eq:W-op}
  \mvc{W}=
\mvc{V}_{\med}^{-1}\,\mvc{U}_{R}^{-1}(h)\,\mvc{V}_{\med}=
\begin{pmatrix}
\mvc{W}_{11} & \mvc{W}_{12}\\
\mvc{W}_{21} & \mvc{W}_{22}
\end{pmatrix}
\end{align}
where $\mvc{U}_{R}(\tau)=\mvc{T}_{\ind{rot}}(-\phi_p) \mvc{U}(\tau,0)
\mvc{T}_{\ind{rot}}(\phi_p)$
is the rotated operator of evolution.
This operator is the solution of the initial value
problem~\eqref{eq:evol_problem} 
with $\mvc{M}(\tau)$ replaced by
$\mvc{M}_R(\tau)=\mvc{T}_{\ind{rot}}(-\phi_p) \mvc{M}(\tau) \mvc{T}_{\ind{rot}}(\phi_p)$.

From Eqs.~\eqref{eq:transm-rel}, and~\eqref{eq:transf-rel},
the transmission and reflection matrices 
can be expressed in terms of 
the transfer matrix as follows  
\begin{align}
&  
\label{eq:W_TR}
\mvc{T}= \mvc{W}_{11}^{-1},
\quad
\mvc{R} =\mvc{W}_{21}\,\mvc{T}.
\end{align}

In what follows we assume 
that, as is illustrated in Fig.~\ref{fig:norm_inc_geom}, 
the light impinges normally onto the CLC cell with 
$q_p=0$ and $\phi_p=0$.
So,  all the waves are propagating along the helical axis 
and we deal with the most studied limiting case of normal incidence, 
which has a long history dating back more than half a century
to the original paper by De Vries~\cite{Dvries:acta:1951}.

\subsection{Operator of evolution: rotating wave ansatz}
\label{subsec:rotating-wave}

By contrast to the case of oblique incidence,
it can be shown that the initial value problem for
the evolution operator~\eqref{eq:evol_problem}
is exactly solvable at $q_p=0$.
To this end, we begin with  
the vector amplitudes written in the circular basis
\begin{align}
  \label{eq:E_circ}
  \vc{E}_{\alpha}=E_x^{(\alpha)}\,\uvc{x}+E_y^{(\alpha)}\,\uvc{y}=E_{+}^{(\alpha)}\,\uvc{e}_{+}
      +E_{-}^{(\alpha)}\,\uvc{e}_{-},
\end{align}
where
$\uvc{e}_{\pm}=(\uvc{x}\pm i\,\uvc{y})/\sqrt{2}$
and
$E_{\pm}^{(\alpha)}=(E_{x}^{(\alpha)}\mp i\, E_{y}^{(\alpha)})/\sqrt{2}$,
so that the transfer and reflection matrices
\begin{align}
  \label{eq:TR_circ}
  \mvc{T}_{\mathrm C}=\mvc{C}\,\mvc{T}\,\hcnj{\mvc{C}},
\quad
  \mvc{R}_{\mathrm C}=\mvc{C}\,\mvc{R}\,\hcnj{\mvc{C}},
\end{align}
where $\displaystyle \mvc{C}=
\dfrac{1}{\sqrt{2}}
\begin{pmatrix}
  1& -i\\1 & i
\end{pmatrix}
$,
relate the circular components of the incident,
transmitted and reflected waves.
When the basis changes
the system~\eqref{eq:matrix-system} transforms 
and, in the circular basis, 
takes the following form 
\begin{align}
&
 \label{eq:system_circ}
  -i\pdrs{\tau}\vc{F}_{\mathrm C}=\mvc{M}_{\mathrm C}\,\vc{F}_{\mathrm C},
\quad
  \vc{M}_{\mathrm C}=\tilde{\mvc{C}}\,\mvc{M}\,\hcnj{\tilde{\mvc{C}}},
\\
&
  \label{eq:F_circ}
  \vc{F}_{\mathrm C}=\tilde{\mvc{C}}\,\vc{F},
\quad
\tilde{\mvc{C}}=
\begin{pmatrix}
  \mvc{C}&\mvc{0}\\ \mvc{0}&\mvc{C}
\end{pmatrix}.
\end{align}

The next step is the rotating wave ansatz
that uses the basis vectors
rotating in helical fashion similarly to the director field.
For the electric field, it can be written
in the following form:
\begin{align}
  \label{eq:E_RW}
  \vc{E}=E_x^{(\ind{rw})}\,\uvc{n}+E_y^{(\ind{rw})}\,\uvc{m}=
E_{+}^{(\ind{rw})}\,\uvc{e}_{+}^{(\ind{rw})}
      +E_{-}^{(\ind{rw})}\,\uvc{e}_{-}^{(\ind{rw})}
\end{align}
where 
$
\uvc{e}_{\pm}^{(\ind{rw})}=
\exp\{\mp i\phi\}
\uvc{e}_{\pm}=(\uvc{n}\pm i\,\uvc{m})/\sqrt{2}
$
and the unit vector
$\uvc{m}=\uvc{z}\times\uvc{n}=\pdrs{\phi}\uvc{n}$
is perpendicular to
the director $\uvc{n}$ defined in Eq.~\eqref{eq:director}.
More generally, this ansatz is defined as follows
\begin{align}
&
  \label{eq:F_RW}
\vc{F}_{\rw}=
\mvc{R}_{+}(\phi)
\, \vc{F}_{\mathrm C},
\\
&
 \label{eq:Rpm}
\mvc{R}_{\pm}(\phi)=
\begin{pmatrix}
\exp\{i\phi\,\bs{\sigma}_3\}&\mvc{0}\\ 
\mvc{0}&\exp\{\pm i\phi\,\bs{\sigma}_3\},
\end{pmatrix}
\end{align}
so that Eq.~\eqref{eq:system_circ}
is transformed into the system
\begin{align}
&
  \label{eq:system_RW}
  -i\pdrs{\tau}\vc{F}_{\rw}=\mvc{M}_{\rw}\,\vc{F}_{\rw}=
\notag
\\
&
    \begin{pmatrix} \tilde{q}\,\bs{\sigma}_3 &\mu\, \mvc{I}_{2}\\
\epsilon_c
\bigl\{
\mvc{I}_2+ u_a\, \bs{\sigma}_{1}
\bigr\}
&\tilde{q}\,\bs{\sigma}_3 \end{pmatrix}
    \begin{pmatrix}\vc{E}_{\rw}\\\vc{H}_{\rw} \end{pmatrix},
\end{align}
where the matrix $\mvc{M}_{\rw}$ is independent
of $\tau$.

The evolution operator
of the system~\eqref{eq:system_RW}
then can be readily expressed in terms of
eigenvalues and eigenvectors
of the matrix $\mvc{M}_{\rw}$.
The result is given by
\begin{align}
&
  \label{eq:matrix-U_rw}
     \mvc{U}_{\rw}(\tau)=\exp\{i \mvc{M}_{\rw}\, \tau\}=
\notag
\\
&
\mvc{V}\,
      \begin{pmatrix}
        \mvc{U}_{+}(n_c \tau)&\mvc{0}\\
\mvc{0}&\mvc{U}_{-}(n_c \tau)
      \end{pmatrix}
\mvc{V}^{-1},
\:
n_c^2=\mu \epsilon_c,
\\
&
  \label{eq:matrix-Ud}
\mvc{U}_{\pm}(\tau)
=
\exp\{\pm i \bs{\Lambda}\, \tau\},
\quad
\bs{\Lambda}
=
\begin{pmatrix}
  \kappa_{1} &0\\
0& \kappa_{2} 
\end{pmatrix},
\end{align}
where 
\begin{align}
&
\label{eq:kappa_i}
\kappa_{1,\,2}=
\left[
1+q_c^2\pm
\sqrt{
4 q_c^2+u_a^2
}
\right]^{1/2},
\quad
q_c=\tilde{q}/n_c,
\\
&
  \label{eq:matrix-V}
  \mvc{V}
=
\begin{pmatrix}
  \mvc{E} & -\bs{\sigma}_{1}\mvc{E}\\
  \mvc{H} & \bs{\sigma}_{1}\mvc{H}
\end{pmatrix},
\:
\mvc{E}=
\bigl(
\vc{E}_1\,\vc{E}_2
\bigr),
\:
\mvc{H}=
\bigl(
\vc{H}_1\,\vc{H}_2
\bigr)
\\
&
\label{eq:EH_i}
\vc{E}_i=
\begin{pmatrix}
  u_a \\
(\kappa_i-q_c)^2-1
\end{pmatrix},
\:
\vc{H}_i=
\frac{n_c}{\mu}\,
\left[\,
\kappa_i\,\mvc{I}_2- q_c\,
\bs{\sigma}_3
\right]\,\vc{E}_i.
\end{align}
Note that the eigenvector matrix~\eqref{eq:matrix-V} 
satisfies orthogonality conditions
of the form~\cite{Kiselev:pra:2008}
\begin{align}
&
  \label{eq:inv_V}
\mvc{V}^{-1}=
  \mvc{N}^{-1}\,\tcnj{\mvc{V}}\,\mvc{G},
\\
&
  \label{eq:matrix-N}
\mvc{N}=\diag(\mvc{N}_{+},-\mvc{N}_{+}),
\quad
\mvc{N}_{+}=\diag(N_1,N_2),
\\
&
N_i=
\frac{2 n_c}{\mu}
\bigl\{
(\kappa_i-q_c) u_a^2+
(\kappa_i+q_c)
[(\kappa_i-q_c)^2-1]^2
\bigr\},
\end{align}
and one of the eigenvalues~\eqref{eq:kappa_i},
$\kappa_2$, is imaginary
in the optical stop band (photonic bandgap):
\begin{align}
  \label{eq:q_pm}
  \kappa_2=i |\kappa_2|,\quad
q_{-}\equiv\sqrt{1-|u_a|}\le
|q_c|\le
q_{+}\equiv\sqrt{1+|u_a|},
\end{align}
where the corresponding eigenmode
becomes evanescent and selective
reflection takes place.
In the photonic bandgap,
additional analysis is required
so as to deal with the problem of
numerical instability caused
by the presence of exponentially large terms
proportional to $\exp(|\kappa_2| h_c)$.
This analysis is given in Sec.~\ref{subsec:photonic-band-gap}. 

We can now write down the resulting expression 
for the evolution operator of the system~\eqref{eq:F_circ}:
\begin{align}
&
  \label{eq:matrix_U_norm}
   \mvc{U}_{\ind{C}}(\tau)=
   \mvc{R}_{+}(-\phi)\,
     \mvc{U}_{\rw}(\tau)\,\mvc{R}_{+}(\phi_{0}).
\end{align}

\subsection{Transmission and reflection matrices}
\label{subsec:transm-refl}

In the case of normal incidence with $q_p=0$ and $\phi_p=0$,
the eigenvector matrix for the ambient medium
in the circular basis 
and the corresponding orthogonality relation
are given by
\begin{align}
&
  \label{eq:Vm_norm}
  \mvc{V}_{\med}=
  \begin{pmatrix}
    \mvc{I}_2 & -\bs{\sigma}_1\\
\dfrac{n_{\med}}{\mu_{\med}}\, \mvc{I}_2 &
\dfrac{n_{\med}}{\mu_{\med}}\, \bs{\sigma}_1
  \end{pmatrix},
\:
\mvc{V}_{\med}^{-1}=
  \mvc{N}_{\med}^{-1}\,\tcnj{\mvc{V}_{\med}}\,\mvc{G},
\\
&
\mvc{N}_{\med}= N_{\med} \mvc{G}_3,
\:
\mvc{G}_3=\diag(\mvc{I}_{2},-\mvc{I}_{2}),
\:
N_{\med}=2 n_{\med}.
\end{align}
For the evolution operator~\eqref{eq:matrix_U_norm}, 
these formulas and the relation
\begin{align}
  \label{eq:Vm_Rpm}
\mvc{R}_{+}(\phi)\,
  \mvc{V}_{\med}= \mvc{V}_{\med}\, \mvc{R}_{-}(\phi)
\end{align}
can now be used to deduce
the transfer matrix~\eqref{eq:W-op}
in the following form:
\begin{align}
&
  \label{eq:W_norm_inc}
  \mvc{W}=\mvc{R}_{-}(-\phi_0)\,\mvc{W}_{\rw}\,\mvc{R}_{-}(\phi_1),
\: \phi_1=\phi_0+\pi\, \nu
\\
&
\label{eq:matrix-W_rw}
     \mvc{W}_{\rw}=\mvc{V}_2\,
      \begin{pmatrix}
        \mvc{U}_{-}(h_c)&\mvc{0}\\
\mvc{0}&\mvc{U}_{+}(h_c)
      \end{pmatrix}
\mvc{V}_{2}^{-1},\: h_{c}=n_{c} h,
\\
&
 \label{eq:V_2}
  N_{\med} \mvc{V}_2=
  \begin{pmatrix}
    \mvc{A}_{+} & \mvc{A}_{-}\\
    \mvc{A}_{-} & \mvc{A}_{+}
  \end{pmatrix},
\:
  \tilde{\mvc{N}}_{+}\, \mvc{V}_2^{-1}=
  \begin{pmatrix}
    \tcnj{\mvc{A}}_{+} & -\tcnj{\mvc{A}}_{-}\\
    -\tcnj{\mvc{A}}_{-} & \tcnj{\mvc{A}}_{+}
  \end{pmatrix},
\end{align}
where $\nu= 2 D/P=q_c h_c/\pi$
is the CLC half-turn number parameter;
$\mvc{V}_2\equiv\mvc{V}_{\med}^{-1}\,\mvc{V}$
and $ \tilde{\mvc{N}}_{+}\equiv\diag(\mvc{N}_{+},\mvc{N}_{+})$.
The matrices $\mvc{A}_{+}$ and $\mvc{A}_{-}$
are given by
\begin{align}
&
\label{eq:A_pm}
\mvc{A}_{+}=\frac{n_{\med}}{\mu_{\med}}\,\mvc{E}+\mvc{H}=
\bigl(\,\vc{a}_{1}^{(+)}\, \vc{a}_{2}^{(+)}\,\bigr),
\notag
\\
&
\mvc{A}_{-}=
\bs{\sigma}_1\,
\Bigl\{
-\frac{n_{\med}}{\mu_{\med}}\,\mvc{E}+\mvc{H}
\Bigr\}=
\bigl(\,\vc{a}_{1}^{(-)}\, \vc{a}_{2}^{(-)}\,\bigr)
\end{align}
and define the block $2\times 2$ matrices, $\mvc{W}_{ij}^{(\ind{rw})}$,
of the transfer matrix~\eqref{eq:matrix-W_rw}
as follows
\begin{subequations}
\label{eq:W_rw_ij}
\begin{align}
  &
  \label{eq:W_rw_11}
  N_{\med} \mvc{W}_{11}^{(\ind{rw})}=
\mvc{A}_{+}
\,
\mvc{W}_{-}
\,
\tcnj{\mvc{A}}_{+}
-
\mvc{A}_{-}
\,
\mvc{W}_{+}
\,
\tcnj{\mvc{A}}_{-},
\\
&
  \label{eq:W_rw_21}
  N_{\med} \mvc{W}_{21}^{(\ind{rw})}=
\mvc{A}_{-}
\,
\mvc{W}_{-}
\,
\tcnj{\mvc{A}}_{+}
-
\mvc{A}_{+}
\,
\mvc{W}_{+}
\,
\tcnj{\mvc{A}}_{-}=
\notag
\\
&
- N_{\med} \tcnj{[\mvc{W}_{12}^{(\ind{rw})}]},
\\
&
  \label{eq:W_rw_22}
  N_{\med} \mvc{W}_{22}^{(\ind{rw})}=
\mvc{A}_{+}
\,
\mvc{W}_{+}
\,
\tcnj{\mvc{A}}_{+}
-
\mvc{A}_{-}
\,
\mvc{W}_{-}
\,
\tcnj{\mvc{A}}_{-},
\end{align}
\end{subequations}
where
\begin{align}
&
  \label{eq:W_pm}
  \mvc{W}_{\mp}=\mvc{U}_{\mp}(h_c)\,\mvc{N}_{+}^{-1}
=
\begin{pmatrix}
  \gamma_{\pm 1}/N_1& 0\\
0 & \gamma_{\pm 2}/N_2
\end{pmatrix}
,
\notag
\\
&
\gamma_{\pm i}=\exp(\mp i \kappa_i h_c).
\end{align}

Finally, for the transmission and reflection matrices~\eqref{eq:W_TR}
in the circular basis,
we have the relations
\begin{subequations}
\label{eq:TR_norm_inc}
\begin{align}
&
  \label{eq:T_norm_inc}
\mvc{T}_{\mathrm C}=
  \exp[-i\phi_1\,\bs{\sigma}_3]
\,
\mvc{T}_{\rw}
\,
\exp[i\phi_0\,\bs{\sigma}_3],
\\
&
 \label{eq:R_norm_inc}
\mvc{R}_{\mathrm C}=
  \exp[i\phi_0\,\bs{\sigma}_3]
\,
\mvc{R}_{\rw}
\,
\exp[i\phi_0\,\bs{\sigma}_3],
\\
&
  \label{eq:TR_rw}
  \mvc{T}_{\rw}=[\mvc{W}_{11}^{(\ind{rw})}]^{-1},
\quad
  \mvc{R}_{\rw}=\mvc{W}_{21}^{(\ind{rw})}\,\mvc{T}_{\rw},
\end{align}
\end{subequations}
where
the block $2\times 2$ matrices
are given in Eqs.~\eqref{eq:W_rw_11} and~\eqref{eq:W_rw_21}.

Note that
 the theoretical curves
presented in Fig.~\ref{fig:ellipt_azim} 
are computed for 
the ellipticity
\begin{align}
  \label{eq:ellipticity}
  \epsilon_{\ind{ell}}=
\frac{|E_{+}^{(\ind{trm})}|-|E_{-}^{(\ind{trm})}|}{|E_{+}^{(\ind{trm})}|+|E_{-}^{(\ind{trm})}|}
\end{align}
of the transmitted wave
\begin{align}
&
  \label{eq:transm_wave}
\exp[i\phi_1\,\bs{\sigma}_3]
  \begin{pmatrix}
    E_{+}^{(\ind{trm})}\\
    E_{-}^{(\ind{trm})}
  \end{pmatrix}
=
\notag
\\
&
\mvc{T}_{\rw}
\,
\exp[i\Delta\phi\,\bs{\sigma}_3]
\begin{pmatrix}
    1\\
    1
  \end{pmatrix}
E_0^{(\ind{inc})},
\:
\Delta\phi=\phi_0-\phi_p^{(\ind{inc})},
\end{align}
where $E_0^{(\ind{inc})}$ and $\phi_p^{(\ind{inc})}$
are the amplitude and polarization azimuth
of the linearly polarized incident wave, respectively.

 \begin{figure*}[!tbh]
 \centering
    \resizebox{130mm}{!}{\includegraphics*{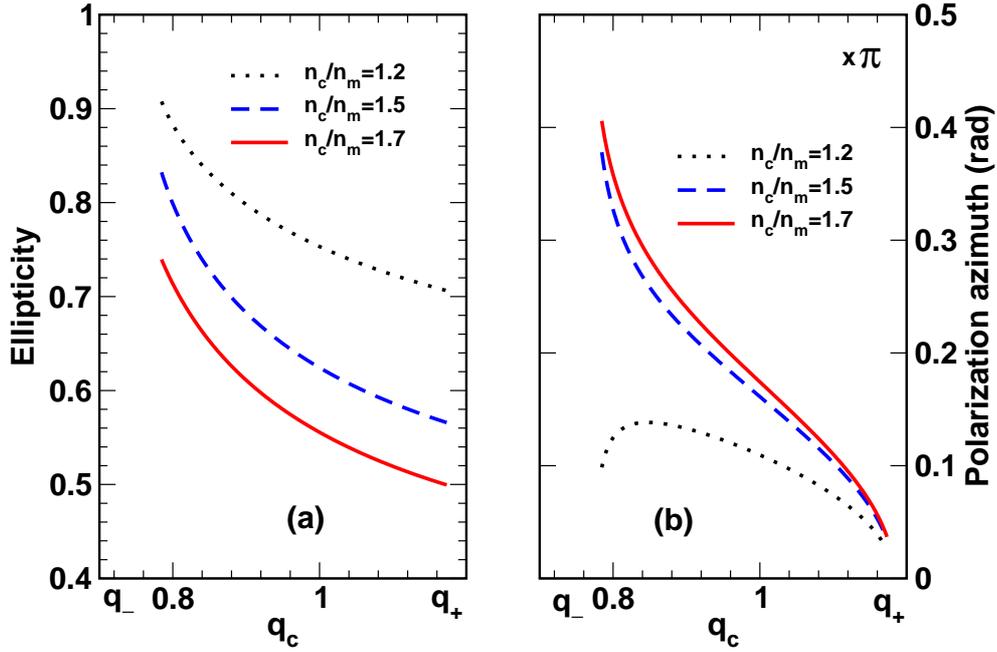}}
 \caption{%
(Color online)
 Ellipticity and polarization azimuthal angle
 of light transmitted through a thick CLC layer
 as a function of the parameter $q_c$ at different values of the optical contrast
 $n_c/n_{\med}$.
 }
 \label{fig:pol_thick}
 \end{figure*}

 \subsection{Analytical treatment in photonic band gap}
 \label{subsec:photonic-band-gap}

 In the photonic band gap (see Eq.~\eqref{eq:q_pm}), 
 the eigenvalue parameter
 \begin{align}
   \label{eq:gamma_def}
   \gamma\equiv \gamma_{+2}=\exp(|\kappa_2| h_c)
 \end{align}
 is large. Since formulas~\eqref{eq:TR_rw} 
 for the transmission and reflection matrices
 contain both large and small terms
 proportional to $\gamma$ and $\gamma^{-1}$,
 they
 cannot be directly applied for numerical analysis.

 In this section we derive the analytical results applicable
 in the optical stop band.
 For this purpose,
 we shall use the dyadic representation for 
 the transfer matrix~\eqref{eq:matrix-W_rw}
 \begin{align}
 &
   \label{eq:dyadic-W_rw}
   \mvc{W}_{\rw}=\sum_{\alpha =  \{\pm 1,\,,\pm 2\}}\gamma_{\alpha}\,
 \vc{v}_{\alpha}\otimes\vc{u}_{\alpha}=
 \notag
 \\
 &
 \gamma \vc{v}\otimes\vc{u}+\bs{\Gamma},
 \quad
 \bs{\Gamma}=
 \sum_{\alpha' \ne +2}\gamma_{\alpha'}\,
 \vc{v}_{\alpha'}\otimes\vc{u}_{\alpha'},
 \end{align}
 where the vectors
 \begin{align}
   \label{eq:vu_pmi}
   \vc{v}_{\pm i}=
   \begin{pmatrix}
     \mvc{P}_1[\vc{v}_{\pm i}]\\
 \mvc{P}_2[\vc{v}_{\pm i}]
   \end{pmatrix}
 =
 N_{\med}^{-1}
   \begin{pmatrix}
     \vc{a}_i^{(\pm)}\\
     \vc{a}_i^{(\mp)}
   \end{pmatrix},
 \quad
   \vc{u}_{\pm i}=
   \begin{pmatrix}
     \mvc{P}_1[\vc{u}_{\pm i}]\\
 \mvc{P}_2[\vc{u}_{\pm i}]
   \end{pmatrix}
 =\pm N_{i}^{-1}
   \begin{pmatrix}
     \vc{a}_i^{(\pm)}\\
     - \vc{a}_i^{(\mp)}
   \end{pmatrix}
 \end{align}
 are expressed in terms of the vector-columns,
 $\vc{a}_{i}^{(\pm)}$,
 given in Eq.~\eqref{eq:A_pm} and
 form a biorthogonal set:
 $\sca{\vc{v}_{\alpha}}{\vc{u}_{\alpha'}}=\delta_{\alpha\,\alpha'}$.
 The latter follows because 
 columns of the matrix $\mvc{V}_2$ and rows of the inverse matrix
 $\mvc{V}_2^{-1}$
 (see Eq.~\eqref{eq:V_2})
 give the components of
 $\vc{v}_{\alpha}$ and $\vc{u}_{\alpha}$, respectively.
 From Eqs.~\eqref{eq:dyadic-W_rw} and~\eqref{eq:vu_pmi}, 
 the block $2\times 2$ matrices can be similarly rewritten in the dyadic form:
 \begin{align}
 &
 \label{eq:W_ij-dyad}
  N_{\med} N_{2}\mvc{W}_{ij}^{(\ind{rw})}=
  N_{\med} N_{2}\sum_{\alpha}\gamma_{\alpha}
 \mvc{P}_i[\vc{v}_{\alpha}]\otimes\mvc{P}_j[\vc{u}_{\alpha}]
 =
 \gamma\mvc{A}_{ij} +\bs{\Gamma}_{ij}.
 \end{align}

 The transmission matrix~\eqref{eq:TR_rw}
 is expressed in terms of $\mvc{W}_{11}^{(\ind{rw})}$
 which is defined by the matrices
 \begin{align}
 &
 \label{eq:A_11-dyad}
 \mvc{A}_{11}=\vc{a}_2^{(+)}\otimes\vc{a}_2^{(+)},
 \\
 &
 \label{eq:G_11-dyad}
 \bs{\Gamma}_{11}=
 N_2/N_1
 \sum_{s= \pm} s \gamma_{s 1}\,
 \vc{a}_{1}^{(s)}\otimes\vc{a}_{1}^{(s)}
 +
 \gamma^{-1} \vc{a}_2^{(-)}\otimes\vc{a}_2^{(-)}
 \end{align}
 that enter the right hand side of
 Eq.~\eqref{eq:W_ij-dyad}.

 Our task is to derive analytical expression for 
 the transmission matrix that does not contain 
 large terms proportional to $\gamma$.
 To this end, we shall use the relations for $2\times 2$ matrices
 \begin{align}
 &
   \label{eq:inverse_gen}
   \mvc{A}^{-1}=[\det\mvc{A}]^{-1}\mvc{A}^{\perp},
 \quad
 \mvc{A}^{\perp}=\bs{\sigma}_2\cdot\tcnj{\mvc{A}}\cdot\bs{\sigma}_2,
 \\
 &
 \label{eq:det_W11}
   |\gamma\mvc{A}_{11}+\bs{\Gamma}_{11}|=
 \gamma\bigl[
 |\mvc{A}_{11}+\bs{\Gamma}_{11}|-(1-\gamma^{-1})
 |\bs{\Gamma}_{11}|
 \bigr],
 \end{align}
 where 
 $|\mvc{A}|\equiv\det(\mvc{A})$
 is the determinant of a matrix $\mvc{A}$ and
 $\mvc{A}^{\perp}\equiv\adj(\mvc{A})$ is the adjugate of a
 $2\times 2$ matrix $\mvc{A}$.
 It is not difficult to see that, for two dimensional vectors, the adjugate of a dyadic
 can be written in the following form
 \begin{align}
   \label{eq:dyad_perp}
  (\vc{x}\otimes\vc{y})^{\perp}=\vc{y}^{\perp}\otimes\vc{x}^{\perp},
 \end{align}
 where $\vc{x}^{\perp}=-i\bs{\sigma}_2\cdot \vc{x}$
 and $\sca{\vc{x}^{\perp}}{\vc{x}}=0$. 

 The transmission matrix~\eqref{eq:TR_rw}
 can now be cast into the form
 suitable for using in the photonic band gap
 as follows
 \begin{subequations}
 \label{eq:T_rw_ph-gap}
 \begin{align}
 &
   \label{eq:T_rw_dyad}
   \mvc{T}_{\rw}=
 \frac{N_{\med} N_{2}}{|\mvc{A}_{11}+\bs{\Gamma}_{11}|-(1-\gamma^{-1})
 |\bs{\Gamma}_{11}|}\,
 \left\{
 \mvc{A}_{11}^{\perp}+\gamma^{-1}\bs{\Gamma}_{11}^{\perp}
 \right\},
 \\
 &
 \label{eq:A11_perp}
 \mvc{A}_{11}^{\perp}=\vc{b}_2^{(+)}\otimes\vc{b}_2^{(+)},
 \\
 &
 \label{eq:G11-perp}
 \bs{\Gamma}_{11}^{\perp}=
 N_2/N_1
 \sum_{s= \pm} s \gamma_{s 1}\,
 \vc{b}_{1}^{(s)}\otimes\vc{b}_{1}^{(s)}
 +
 \gamma^{-1} \vc{b}_2^{(-)}\otimes\vc{b}_2^{(-)},
 \\
 &
 \label{eq:b_i}
 \vc{b}_i^{(s)}=-i\bs{\sigma}_2\cdot\vc{a}_i^{(s)}
 =
 \begin{pmatrix}
   0 & -1\\
 1 & 0
 \end{pmatrix}
 \vc{a}_i^{(s)}.
 \end{align}
 \end{subequations}

 From Eq.~\eqref{eq:T_rw_dyad},
 it is clear that in the limit of thick cells
 where $\gamma\to 0$
 the transmission matrix
 approaches the singular dyadic
 \begin{align}
   \label{eq:T_approx}
   \mvc{T}_{\rw}\propto
 \vc{b}_2^{(+)}\otimes\vc{b}_2^{(+)},
 \quad
 \gamma^{-1}\to 0
 \end{align}
 defined by the vector $\vc{b}_2^{(+)}$.
 The polarization characteristics of this vector
 are plotted in Fig.~\ref{fig:pol_thick}
 as a function of $q_c$ at different values
 of the contrast ratio $n_c/n_{\ind{m}}$.
 For any polarization state of the incident light
 with $\vc{E}_{\ind{inc}}\nparallel \exp[-i\phi_0\,\bs{\sigma}_3]\,\vc{a}_2^{(+)}$,
 the parameters 
 depicted in Fig.~\ref{fig:pol_thick}
 determine the ellipticity and the polarization azimuth
 of the transmitted wave
 when the CLC cell is sufficiently thick.
 Since $\sca{\vc{a}_2^{(+)}}{\vc{b}_2^{(+)}}=0$,
 transmission of the incident wave with  
 $\vc{E}_{\ind{inc}}\parallel
 \exp[-i\phi_0\,\bs{\sigma}_3]\,\vc{a}_2^{(+)}$
 has been completely suppressed
 in the thick cell limit $\gamma\to 0$.

 Equation~\eqref{eq:TR_rw} gives the reflection
 matrix expressed in terms of the transmission
 matrix~\eqref{eq:T_rw_dyad}
 and $\mvc{W}_{21}^{(\ind{rw})}$.
 The latter is defined in Eq.~\eqref{eq:W_ij-dyad}
 with the matrices given by
 \begin{align}
 &
 \label{eq:A_21-dyad}
 \mvc{A}_{21}=\vc{a}_2^{(-)}\otimes\vc{a}_2^{(+)},
 \\
 &
 \label{eq:G_21-dyad}
 \bs{\Gamma}_{21}=
 -N_2/N_1
 \sum_{s= \pm} s \gamma_{s 1}\,
 \vc{a}_{1}^{(-s)}\otimes\vc{a}_{1}^{(s)}
 +
 \gamma^{-1} \vc{a}_2^{(+)}\otimes\vc{a}_2^{(-)}.
 \end{align}
 Similar to the transmission matrix,
 the reflection matrix  can now be written in the following dyadic form: 
 \begin{align}
   \label{eq:R_rw_dyad}
   \mvc{R}_{\rw}=
 \frac{\bs{\Gamma}_{21}\cdot\mvc{A}_{11}^{\perp}+
 (\mvc{A}_{21}+
 \gamma^{-1}\bs{\Gamma}_{21})\cdot\bs{\Gamma}_{11}^{\perp}}
 {|\mvc{A}_{11}+\bs{\Gamma}_{11}|-(1-\gamma^{-1})
 |\bs{\Gamma}_{11}|},
 \end{align}
 where the orthogonality relation $\mvc{A}_{12}\mvc{A}_{11}^{\perp}=\mvc{0}$
 is taken into account.
 From Eq.~\eqref{eq:R_rw_dyad}, it can be seen
 that, by contrast to the transmission matrix~\eqref{eq:T_rw_dyad},
 the reflection matrix is non-singular in the limiting case of thick
 CLC cells. 

 We conclude this section with the remark that
 formulas~\eqref{eq:T_rw_dyad}
 and~\eqref{eq:R_rw_dyad} are exact and remain applicable 
 outside the photonic band gap.


\end{document}